\documentstyle[12pt,psfig]{article}
\def\bi{\bibitem}
\newcommand{\be}{\begin{equation}}
\newcommand{\ee}{\end{equation}}
\newcommand{\beq}{\begin{eqnarray}}
\newcommand{\eeq}{\end{eqnarray}}
\newcommand{\bear}{\begin{array}}
\newcommand{\ear}{\end{array}}

\begin{document}
\title{ The CWKB approach to non-reflecting potential\\
and cosmological implications.}
\author{S. Biswas$^{*}$ and I. Chowdhury  \\Department of Physics, University of Kalyani, West Bengal,\\
India, Pin.- 741235\\
$*$email:sbiswas@klyuniv.ernet.in}\date{}\maketitle
\begin{abstract} We discuss the method of calculating the reflection coefficient using complex trajectory WKB (CWKB) 
approximation. This enables us to give an interpretation of non-reflecting nature of the potential under certain conditions 
and clarify some points, reported incorrectly elsewhere \cite{vs:ejp} for the potential $U(x)=-U_0cosh^2(x/a)$. We show that the repeated reflectios between the turning points are essential, which most authors overlooked, in obtaining the non-reflecting c
ondition. We find that the considered repeated reflection paths are in conformity with Bogolubov transformation technique. We discuss the implications of the results when applied to the particle production scenario, considering $x$ as a time variable and 
 also stress the cosmological implications of the result with reference to radiation domonated and de Sitter spacetime.
\end{abstract}
\section{Introduction}
In a communication Vugalter and Sorokin \cite{vs:ejp}  studied the problem of reflection of a particle from the potential well
\be
U(x)=-U_0cosh^2(x/a)
\ee
using the WKB approximation as described by Landau \cite{ll:qm} for systems having complex turning points. The turning points obtained in 
 \cite{vs:ejp}  are wrong and their explanation for introducing the extra phase at one of the turning points seems unsatisfactory, though they 
obtained accidently the correct result. We obtain here the correct result using the complex trajectory WKB approximation. The CWKB
 method proposed and developed by Biswas \cite{bis1:pjp, bis2:pjp, bis3:cqg, bis4:cqg, bis5:grg, bis6:ijmp} has been tested in many 
applications. We apply this method for the potential given in (1) to obtain the reflection coefficient. It is found that when we consider repeated reflections between the turning points, the correct non-reflecting condition is obtained. We choose the pote
ntial (1) because for this case the exact result is known through Bogolubov technique. To apply the above potential to cosmological particle production we have to replace $x$ by conformal time $\eta$ where $\eta$ is the conformal time in FRW spacetime. In
 that case the scale factor is given by $a^2(t)=\frac{2U_0}{m}+\frac{t^2}{a^2}$ and represents an expanding spacetime. The reflection coefficient in such a spacetime will then represent pair production amplitude. One characteristic feature of the spacetim
e is that asymptotically we have $da(t)/dt= \rm costant$ so that the number of particles will show a Planckian spectrum. This provides the particle production as an initiator to decoherence as if we have ` decoherence without decoherence '. We will also d
iscuss this aspect in this article.
\par
The motivation behind studying (1) is to find correct CWKB paths in conformity with Bogolubov transformation technique. This helps us switch over to cosmological applications in which the CWKB paths play an important role in classicalization of quantum fl
uctuations that act as seeds of density perturbations. The classical nature of fluctuations is usually assumed, or argued for in passing, by invoking Gibbons-Hawking radiation or the horizon in de Sitter space. Here we will find that horizon or coordinate
 dependent horizons (as in the present example) lead to particle production in the same way in CWKB. This result has a deep significance and is related to the entropy of spacetime \cite{pd:mpl}. In this work we concentrate mainly on the implications of no
n-reflecting condition with respect to expanding spacetime.
\par
In section 1 we mention the main result of non-reflecting potential (1) obtained from Bogolubov technique. In section 2 we calculate the reflection coefficient using CWKB method. Cosmological implications of the results are discussed in section 3. We end 
up with a concluding discussion.
For the Schr\"{o}dinger equation 
\be
-\frac{\hbar^2}{2m}\frac{d^2\psi}{dx^2}+U(x)\psi=E\psi
\ee
with the potential (1), the reflection amplitude has been calculated \cite{fl:pqm}  with the result 
\be
\mid R\mid ^2=\frac{cos^2\left(\frac{\pi}{2}\sqrt{1+8mU_0a^2/ \hbar^2} \right)}{sinh^2(\pi k_0a)+cos^2\left(\frac{\pi}{2}\sqrt{1+8mU_0a^2/ \hbar^2 }\right)}
\ee
where $k_0=\sqrt{2mE}/ \hbar.$  It now follows from (3) that for $E\,\, \rm and\,\, U_0$ both positive or negative, the reflection coefficient 
$R$ is zero if the condition 
\be
mU_0a^2/ \hbar^2=n(n+1)/2,\,\,\,n=1, 2, 3, ...
\ee
is satisfied. Under this condition the potential is said to be non-reflecting.
\par
 The equation (2) when cosidered  as a potential problem 
has  no turning point at any real $x$ but $U(x) $ treated as complex function may have turning points at complex values of $x$. In that case 
the complex WKB trajectory can turn from the complex turning points and emerge into the real plane to give the reflection component. 
In other words the wave function at any real point ( obtained as a sum over semclassical WKB paths ) is not only contributed by real 
WKB trajectory but also gets contribution from complex WKB paths. We will discuss in this paper the method of CWKB and
 obtain the result (3)  We will restrict our discussion to the case when eqn. (2) has two complex turning points.
\section{Reflection coefficient in the CWKB approximation }
We have given in \cite{bis1:pjp, bis2:pjp, bis3:cqg, bis4:cqg, bis5:grg, bis6:ijmp} the detailed derivation of the reflection coefficient. However 
here we give the basics of the CWKB method  for the case having two complex turning points. We write (2) in the form
\be
\frac{d^2\psi}{dx^2}+\omega^2(x)\psi=0
\ee
where
\be
\omega(x)= \sqrt{2m(E-U)/ \hbar^2}
\ee
Suppose we have two complex turning points $x_1\,\,\rm and\,\,x_2$. Hence
\be
\omega(x_{1,2})=0
\ee
Let us calculate the the wave function at a real point $x$. The contribution at $x$ comes from two types of trajectories, namely, the
direct trajectory (DT)  and the reflected trajectory (RT). The direct trajectory starts from a real point $x_0>>x$ and is at the right of the potential 
barrier. Its contribution in the WKB approximation is written as
\be
\rm DT\simeq \frac{1}{\sqrt \omega(x)}  exp\left(+i\int_{x_0}^x \omega(x^\prime)dx^\prime\right)
\ee
Here $x_0>>x$ lies to the right of $x$. The trajectory that starts from $x_0$ and reaches $x$ is called the direct trajectory. For\,\, reflected\, trajectory\, we\, will\, get\, two\, types\, of contributions\,\,\, {\bf ( I )} $\rm {RT}_{1),2)}$ and { \bf
 ( II )} $\rm {RT}_{1,2)}.$\\
{\bf{Case I:}} $\rm {RT}_{1),2)}$ is the reflection coefficient for the case when the reflection occurs from the two turning points separately.
\par
 Here the trajectory starts from $x_0>>x$ and moving leftward arrive at the turning point $x_1$, turning around $x_1$  
moves back to the real point $x$. Its contribution is
\beq
\rm {RT}_{1)} & \simeq  & (-i) \frac{1 } {\sqrt {\omega(x)}} exp\left(+i\int_{x_0}^{x_1}  \omega(x^\prime)dx^\prime\right) exp\left(-i\int_{x_1}^x
 \omega(x^\prime)dx^\prime\right) \nonumber\\
 & \simeq  &    (-i) \frac{ 1 } {\sqrt {\omega(x)}}  exp\left(+2i\int_{x_0}^{x_1}  \omega(x^\prime)dx^\prime\right) exp\left(-i\int_{x_0}^x
 \omega(x^\prime)dx^\prime\right)
\eeq  
Here $(-i)$ has been introduced for reflection at the turning point $x_1$. We will get similar contribution $\rm{RT}_{2)}$ from the 
turning point  $x_2$ and the reflected wave is then given by
\be
\rm{RT}_{1), 2)} = \rm {RT}_{1)} +\rm {RT}_{2)} 
\ee
The coefficient of $\rm {RT}_{1)} +\rm {RT}_{2)} $ i.e., of 
\begin{center}
 $ \frac{1 } {\sqrt {\omega(x)}}exp\left(-i\int_{x_0}^x \omega(x^\prime)dx^\prime\right)$
\end{center}
in (9) gives the reflection amplitude as we have assumed unit amplitude to be incident
 from the right. Thus
\be
R_{1),2)} =-i\left[exp\left(+2i\int_{x_0}^{x_1}  \omega(x^\prime)dx^\prime\right)
+exp\left[+2i\int_{x_0}^{x_2}  \omega(x^\prime)dx^\prime\right)\right]
\ee
Here $R_{1),2)}$ means the reflection amplitude when the reflection occurs separately from the two turning points.\\
 { \bf {Case II:}} $\rm {RT}_{1,2)}$ is the reflection coefficient when the reflection occurs such that reflection coefficient is due to reflection from a single turning point times the repeated reflections from the two turning points. This case is applic
able when the turning points are close whereas the previous case is applicable when the turning points are far apart.
\par 
Here the trajectory starts from $x_0$, reaches $x_1$ and goes to $x_2$, suffers repeated reflection between 
$x_1\,\, \rm and\,\, x_2$ and finally goes to $x$. As before we get
\beq
\rm{RT}_{1,2)} & \simeq & -i exp\left(+2i \int_{x_0}^{x_1}\omega(x^\prime)dx^\prime\right)\sum_{n=0}^{\infty}\left(-iexp(i\int_{x_2}^{x_1} 
\omega(x^\prime)dx^\prime)\right)^{2n}\nonumber\\
 &  & ~~~~~~~\times exp\left(-i\int_{x_0}^x
 \omega(x^\prime)dx^\prime\right) 
 \nonumber\\
& \simeq & -i\frac{ exp\left( +2i \int_{x_0}^{x_1}\omega(x^\prime)dx^\prime \right)} {1+exp\left( +2i\int_{x_2}^{x_1}
 \omega(x^\prime)dx^\prime \right) } exp\left(-i\int_{x_0}^x
 \omega(x^\prime)dx^\prime\right)
\eeq
In (12) the summation has been replaced by the geometric series. Hence the reflection amplitude is
\beq
R_{1,2)} & = & -i\frac{ exp\left( +2i \int_{x_0}^{x_1}\omega(x^\prime)dx^\prime \right)} {1+exp\left(  +2i\int_{x_2}^{x_1}
 \omega(x^\prime)dx^\prime \right) }
\eeq
Here $R_{1,2)}$ denotes the reflection from the turning point $x_1$ times the contribution of repeated reflections between 
the turning points $x_1$ and $x_2$. Let us apply this result for potential problem given in (1)
\par
The complex \,turning\, points\, are\, determined\, from the \,roots of \,the \,equation $\omega(x)=0$ i.e.,
\be
cosh(x/a)=\pm i \sqrt{\frac{U_0}{E}}
\ee
This gives
\be
exp(x_0^\pm/a)=\pm i \sqrt{\frac{U_0}{E}} \pm i \sqrt{1+\frac{U_0}{E}}
\ee
Hence the roots are
\be
x_0^{\pm}=a\left[\pm i\pi
 + ln\left(\sqrt{1+\frac{U_0}{E}}+\sqrt{\frac{U_0}{E}}\right)\right]
\ee 
Now we are to decide between the cases I and case II. Obviously, if the points $x_0^\pm$ are remote from each other such that 
\be
\mid \int_{x_0^-}^ {x_0^+ } \omega(x)dx \mid>>1
\ee
the contribution of repeated reflection goes to zero (see (13)). For such a situation we are to consider the case I ~~i.e.,  
the expresion for $R_{1)2)}$. Henceforth we will denote the reflection amplitude as $R$ for the case I. To calculate $R$ we evaluate
\beq
I(x)=\int^x \omega(x^\prime)dx^\prime & =  & \frac{k_0a}{2}\left[ln\frac{\sqrt{cosh^2(x/a)+U_0/E } + sinh(x/a)}
 {\sqrt{cosh^2(x/a)+U_0/E} - sinh(x/a)} \right] -i\frac {k_0a}{2}\times\nonumber\\
                                                &    &\left[\sqrt{U_0/E}\,\, ln\frac{\sqrt{cosh^2(x/a)+U_0/E } +i \sqrt {U_0/E }\,\, sinh(x/a)}
 {\sqrt{cosh^2(x/a)+U_0/E } - i \sqrt {U_0/E }\,\, sinh(x/a)} \right]\nonumber\\
\eeq        
Now we are to evaluate the above integral at the turning points . Now at any real point $x_1$ both the logarithmic function give real 
contribution, hence in $\mid R \mid$ this part will not contribute. So we will write all the inessential factors in a phase $\delta$. 
At the turning point $x_0^\pm$, the square root term is always zero. At both the turning points the first logarithmic function contribute
 a phase factor $exp(i\pi)$ as it has no branch point at the turning points. Let us look at the behaviour of second logarithmic function 
as we go from $x=Re(x_0^+)$ to $x=x_0^+$. In that case (the second logarithmic function) the argument of the numerator changes 
from $arctan(U_0/{\sqrt{E(E+2U_0)}})$ to $i\pi$ whereas the denominator goes from $-arctan(U_0/{\sqrt{E(E+2U_0)}})$ to $0$. Hence 
the change of phase of the second logarithmic function is $i\pi$. Thus we get
\be
I(x_0^+)=i\frac{k_0a\pi}{2}+\sqrt{U_0/E}\,\pi\frac{k_0a}{2}
\ee   
For the turning point $x_0^ -$, the first logarithmic function gives the phase $i\pi$ as usual but the second lagarihmic function now 
has a different behaviour. At $x=-i\frac{\pi}{2}+ln\{...\}$, the argument of the numerator of the second logarithmic function 
is zero whereas the denominator has the argument $i\pi$. Thus the argument of the second logarithmic function is now $-i\pi$ .
Hence 
\be
I(x_0^-)= i\frac{k_0a\pi}{2} -\sqrt{U_0/E}\,\pi\frac{k_0a}{2}
\ee
Denoting the reflection amplitude $R_{1) 2)}$ by $R_{CWKB}$  we get
\be
R_{CWKB}=-2ie^{i\delta} e^{-\pi k_0 a}cos\left(\pi \sqrt{2mU_0}\,\,a/ \hbar \right)
\ee
All the inessential factors are now introduced in $\delta$. The different behaviour of the second logarithmic function is due to the
 fact that whereas the numerator of the second logarithmic function has branch point at $x=+i\pi /2$, the denominator has branch 
point at $x=-i\pi /2$.  Thus 
\be
\mid R_{CWKB} \mid ^2 =4e^{-2\pi k_0 a}cos^2\left(\pi \sqrt{2mU_0}\,\,a/ \hbar \right)
\ee
We now evaluate (17) and find
\be
\pi \sqrt{2mU_0}\,a/\hbar >>\,1
\ee
If we apply this condition to the expression (13) ( i.e., for the case II ) we get 
\be
\mid R \mid ^2\simeq e^{-2k_0\pi}
\ee
The effect of repeated reflection vanishes. It therefore implies that we are to consider the case I and introduce the repeated reflections, different from the way we take in for case II. We will now discuss how the repeated reflections are incorporated in
 case I.
\par
From this result it is clear that we must take consider higher order paths (i.e., repeated reflections) that contribute equally 
to the reflected part from $x_0^+$ and $x_0^-$. To be transparant we now write 
\be
S(x_f,x_i)=\int_{x_i}^{x_f} \omega(x)dx
\ee
We consider now paths that incorporate double reflection starting at $x_0$ and ending at $x$ as follows. We consider the path
$x_0 \rightarrow x_1 \rightarrow x_2 \rightarrow x_0 \rightarrow x_1 \rightarrow x$. One may visualize the trajectory by connecting paths between $x_0$ and $x$ as mentioned above. Its contribution is (now denoting the turning 
point as $x_1$ and $x_2$)
\beq
C(1) & = & e^{iS(x_1 , x_0) +iS(x_2,x_1)}\times(-i)e^{-iS(x_0,x_2)+iS(x_1,x_0)}\times(-i)e^{-iS(x,x_1)}\nonumber\\
       & = & (-i)e^{2iS(x_1,x_0)}\left[(-i)e^{ -iS(x_2,x_1)+iS(x_2,x_0)+iS(x_1,x_0)}\right]e^{-iS(x,x_0)}
\eeq
introducing a factor $(-i)$ where there is a reflection around the turning point. We will get similar contribution $C(2)$ for the second turning point with $x_1$ replaced by $x_2$. So for the problem we study,
 this contribution will be
\be
C(1)+ C(2)  =  (-i)\left[e^{2iS(x_1,x_0)}+e^{2iS(x_2,x_0)}\right]\left[(-i)e^{ -iS(x_2,x_1)+iS(x_2,x_0)+iS(x_1,x_0)}\right]e^{-iS(x,x_0)}
\ee
This is one loop contribution between $x_1$ and $x_2$. The second bracket in (27) gives the contribution of double reflection at $x_1$ and $x_2$. This comes out to be the same for the two turning points, only in the numerator $x_1$ is replaced by $x_2$. W
hen we take multiple reflections the result is
\be
R_{1,2)}=-i\frac{e^{2iS(x_1,x_0)}+e^{2iS(x_2,x_0)}} {1+ e^{ -iS(x_2,x_1)+iS(x_2,x_0)+iS(x_1,x_0)} }
\ee
If we evaluate this expression we finally get
\be
 R_{1,2)} \simeq 2\frac{e^{i\delta}  e^{-\pi k_0 a} cos(\pi \sqrt{2mU_0}\,a/ \hbar)}
 {1+e^{-2\pi k_0a + 2i(\pi \sqrt{2mU_0}\,a / \hbar)}}
\ee
Hence
\be
\mid R_{1,2)} \mid^2 \simeq \frac{cos^2\left(\pi \sqrt{2mU_0}\,a/ \hbar\right)} {sinh^2(\pi k_0a)+cos^2\left(\pi \sqrt{2mU_0}\,a/ \hbar\right)}
\ee
i.e., the exact result eqn. (3) under the condition $\pi \sqrt{2mU_0}\,a/\hbar >>\,1$. The result (30) indicates that one must also take into account this type repeated reflections, hitherto unnoticed or overlooked by the workers in this field. For this r
eason our technique may be considered as an extension of Landau's technique.
\section{Cosmological application}
We have stressed earlier [3 - 8] that the CWKB technique is well applicable when the variable $x$ is replaced by a time variable. In that case the the reflection amplitude will be pair production amplitude. Consider the expanding space time 
\be
ds^2=a^2(\eta)\left[d\eta^2- (dx^2+dy^2+dz^2)\right]
\ee
with $a(\eta)$ being the scale factor in conformal time. The temporal equation for a scalar field now reads (we now set $\hbar=1$)
\be
\frac{d^2\psi}{d\eta^2}+\omega^2(\eta)\psi=0
\ee
where $\omega^2(\eta)=k^2+m^2\,a^2(\eta)$. Comparing this equation with (5) we find
\be
a^2(\eta)=\frac{2U_0}{m}cosh^2\frac{\eta}{\eta_0}
\ee
with $a$ in (1) being replaced by $\eta_0$.
It is now evident from (29) and (30) that the pair production amplitude is zero if 
\be
m=\frac{1}{2U_0\eta_0^2}(n+\frac{1}{2})^2
\ee
This result seems to be surprising.
On the other hand if we neglect the cosine square term or take $2mU_0\eta_0^2=n^2$, we find that the spectrum of produced particles show a thermal spectrum with
\be
N\simeq exp(-\frac{k}{T})
\ee
with temperature $T=1/{2\pi\,\eta_0}$. The implication of this result is that quantum fluctuations at the early universe somehow due to particle production decohere to give rise to thermalized spectrum. Here we need not to enforce the interaction which is
 deemed essential in coherent studies. Here classicalization occurs due to particular property of spacetime as if we have a thermal bath with temperature $T=1/{2\pi\,\eta_0}$. It is obvious from the expression of $\vert R\vert^2$ with  cosine term that th
ere will be some oscillations even at late times. This oscillations in the thermalized spectrum is interesting from cosmological point of view. The non-reflecting component if be present during evolution i.e., if the masses of particles are $m=\frac{\hbar
^2}{2U_0\eta_0^2}(n+1/2)^2$, these particles will not be produced during the expansion of the unverse. In other words these massive particles will not show themselves in the thermalized spectrum. No doubt this result is important from cosmological point o
f view. Let us try to understand this aspect. If we assume a radiation dominated stage before a pre-inflationary period, some particles (that satisfy (34)) do not undergo the inflationary expansion and hence do not take part in the thermalized spectrum. T
hese particles thereby remain as non-participating components in the large scale structure formation. Obviously these could turn into dark matter component of our unverse. 
\par
This is an example where the non-reflecting type potential might play a decisiverole in the evolution of the unverse.
\section{Discussion}
We note that the cosine in the numerator results from the existence of two points of reflection whereas the cosine term in the denominator 
arises due to repeated reflections between the turning points. The arguments of cosine equals half the phase difference between two waves 
reflected from $x_0^\pm$. So the condition of no reflection is 
\be
mU_0a^2/ \hbar^2= (n+\frac{1}{2})^2/2\,\,\,\, n= 1,\, 2,\, 3\,...
\ee
This example is a very interesting application of CWKB. Now we can give an explanation of the origin of the expressin like (35) in expading spacetime. The region between the two turning points (euclidean sections) acts like a blackbody resonator at temper
ature $T$. Repeated reflections generate the thermal spectrum which leaks out through the turning point in pur real unverse. One may have some apathy to these arguments and disagree but the emergence of temperature $T=H/{2\pi}$ in de Sitter spacetime usin
g the same technique lends support  to our arguments, Ofcourse, we need further investigations. The emergence of oscillations in the produced particle spectrum is also found in de Sitter spacetime for some restricted values of $m/H$ \cite{mm:prd}and is du
e mainly to the CWKB paths, where $\omega^2$ become negative. We will take up this aspect of decoherence through particle production very soon.  \\
{\bf{Acknowledgement}}\\
This work was done during author's (S. B.) stay at the Assam University, Sichar. The author acknowledges the hospitality provided by 
Dr. R. Bhattacharyya and Dr. A. Sen of Physics department,  AUS.  
     
\end{document}